\documentclass[ 
preprint,
showpacs,
preprintnumbers,
amsmath,amssymb,
aip,
floatfix,
]{revtex4-1}

\usepackage{graphicx}
\usepackage{bm}

\begin{document}

\title{The slow collisional \textbf{E}\textbf{x}\textbf{B} ion drift characterized as the major instability mechanism of a poorly magnetized plasma column with an inward-directed radial electric field}

\author{Thi\'{e}ry PIERRE}

\affiliation{Centre National de la Recherche Scientifique, UMR 7345 
Laboratoire PIIM,  Aix*Marseille University, Marseille, France.}


\begin{abstract}
The low-frequency instability of a cylindrical poorly magnetized plasma with an inward-directed radial electric field is studied changing the gas pressure and the ion cyclotron frequency. The unstable frequency always decreases when the gas pressure is increased indicating collisional effects. At a fixed pressure, the unstable frequency increases with the magnetic field when the B-field is low and decreases at larger magnetic field strength. We find that the transition between these two regimes is obtained when the ion cyclotron frequency equals the ion-neutrals collision frequency. This is in agreement with the theory of the slow-ion drift instability induced by the collisional slowing of the electric ion drift (A. Simon, Phys. Fluids 6, {\bf 382}, 1963).

\end{abstract}


\maketitle

\section{INTRODUCTION}
Understanding the physical mechanisms leading to instabilities and anomalous transport in a magnetized plasma has been a major goal in plasma physics during the past sixty years. In general, magnetized plasmas exhibit macroscopic non-MHD instability due to the radial decrease of the density and to the existence of a radial electric field. In the case of a magnetized plasma column with cylindrical geometry, if the physics of the sheath at the end of the plasma is not included, the situation is highly simplified. However, the radial density gradient, the radial profile of the plasma potential and the radial profile of the electron temperature can still lead to many instabilities. Among them, the electrostatic instabilities are the most violent and they dominate the other types, e.g. density gradient instability, temperature gradient instability, and micro-instabilities.

In a previous paper\cite{gravier}, we have shown experimentally that the rotation of the magnetized plasma column obtained in laboratory plasma devices is due to the large radial inward-directed electric field that can trigger flute-like modes, especially at low B-field. The aim of the present work is to carry out the analysis of the evolution of that dominant low-frequency instability when the pressure and the magnetic field strength are varied. The role of the ion-neutral collisions is investigated. We conclude the paper about the physical destabilization mechanism. To the best of our knowledge, the role of the ion-neutral collisions has not been carefully examined yet in the case of the physical system described here.

\section{The magnetized plasma column}

The plasma is produced in the MISTRAL Machine that has been depicted in previous papers.\cite{matsuk,pierre} The device is similar to the linear plasma device MIRABELLE\cite{mirab} that was used in previous investigations. A weakly ionized magnetized plasma column is produced at low magnetization. The device consists in a stainless steel cylindrical vessel (internal diameter D = 40 cm) evacuated to a base pressure of about $10^{-4}$ Pa.
The formation of a sharp boundary between the magnetized plasma and the vacuum is obtained inserting a stainless steel limiter made of a diaphragm with a free diameter of 8 cm inserted at the entrance of the tube covered by the solenoidal coils. The solenoid is made of 20 water-cooled coils equally spaced along the cylindrical vacuum chamber. The maximum magnetic field strength used in this investigation is 25 mT with a low ripple of the field lines along the plasma column. The plasma column is terminated by a glass end-plate. 
A weakly-ionized plasma is created inside the large source chamber (80 cm diameter) by thermionic discharge using 32 tungsten filaments (0.2 mm in diameter, 14 cm in length) located in front of a large multipolar magnetic anode. The filaments are Joule-heated at 2000 K and they emit the energetic ionizing electrons that are injected inside the magnetized column. A very fine tungsten mesh grid (78\% optical transparency) inserted at the entrance section of the column allows an electric insulation between the source plasma and the target plasma. The grid is kept at floating potential in the experiments reported here. Only energetic electrons overcome the potential of the grid (typically -25 volts) and produce the ionization of the column entering the solenoid through the circular aperture. A linear magnetized plasma column is produced (14 cm diameter, 90 cm length) inside the target chamber.

With a floating injection grid and a biasing of the anode inside the source chamber below ground are established, a saturated low-frequency instability is present around the magnetized plasma column. Decreasing the potential of the anode inside the source chamber, the radial electric field across the magnetized plasma column is increased and the frequency of the unstable mode is higher. This phenomenon has been studied and reported in a recent paper.\cite{jaeger} It is easy to explain the role of the biasing of the anode by the fact that the decrease of the potential of the anode inside the source plasma leads to a larger flux of ionizing electrons injected in the target plasma. This in return gives a lower plasma potential inside the magnetized plasma column and as a consequence, the inward-directed radial electric field increases. More precisely, when the anode inside the source chamber is positively biased, no instability is present. Decreasing the potential of the anode to -5 volts leads to the destabilization of the plasma column and a strongly nonlinear regime of the density fluctuations is obtained. As a consequence of the increased electric field, the \textbf{E}\textbf{x}\textbf{B} drift velocity of both ions and electrons is higher. The flute-like character of the instability has been analyzed in previous papers\cite{rebont} with no phase detection along the plasma column. 

\section{Experimental results}

During the measurements described in this paper, the typical B-field strength on the axis is varied between 5 mT and 20 mT. As shown in Fig.1, the radial density profile is almost Gaussian near the axis of the column with a central density in the range $10^{15}$ to $10^{16} m^{-3}$. The electron temperature ranges from 3 to 4 eV depending on the working gas pressure\cite{matsuk}, ranging from 5.$10^{-3}$ Pa to 5.$10^{-2}$ Pa in argon. Several Langmuir probes are used for measuring the radial profile of the plasma density, the plasma potential and the electron temperature. The time-averaged plasma potential profile is roughly parabolic in the central part of the plasma column\cite{barni} corresponding to an inward-directed electric field with linear radial profile. This corresponds to a rigid-body rotation of the core plasma column.

\begin{figure}
	\includegraphics[scale=0.3]{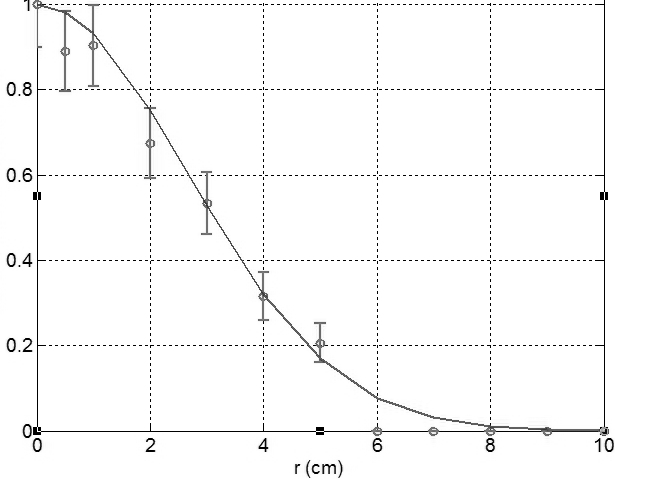}
	\caption{\label{fig:simon-fig1} Mean  radial profile of the electron density.
	}
	\end{figure}

The electron density fluctuations are recorded using cylindrical probes biased at plasma potential and recorded by digital storage oscilloscopes. The power spectrum of the density fluctuations is recorded using an analog spectrum analyzer (HP8560A).\\
We have chosen to set the experiment in the most unstable regime which is obtained using a floating collector. The potential of the anode in the source plasma is 5 volts below the ground. Time series of the density fluctuations are detected by Langmuir probes located at the edge of the central plasma column. The study of the correlation between time series recorded by two probes located on opposite positions at the edge of the plasma column exhibits opposite phase in the signals and allows to determine the m = 1 structure of the mode. The relative fluctuation level is maximum at the edge of the magnetized plasma column and the absolute fluctuation level is maximum at the radial position where the density gradient is maximum. The time series of the density fluctuations detected by the Langmuir probe at the edge of the plasma column is shown in Fig. 2. 
\begin{figure}
	\includegraphics[scale=0.3]{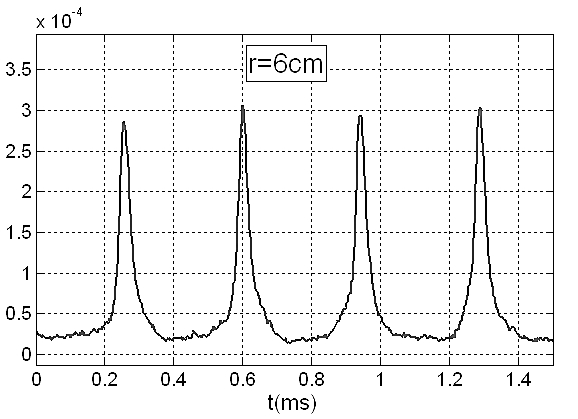}
	\caption{\label{fig:simon-fig2} Time series of the electron density recorded at the edge of the magnetized plasma column in typical conditions exhibiting a strong nonlinear modulation .
	}
	\end{figure}
    
The shape of the signal is similar to a cnoidal wave that can be considered as an infinite sum of periodically repeated solitary waves.\cite{cnoidal} This indicates the strongly nonlinear saturated state of the instability. It is confirmed by the spectral analysis exhibiting multiple harmonics of the unstable frequency. In this situation, the instability is established at a resonant azimuthal wave number. Changing the plasma parameters, the evolution of the unstable frequency is investigated in order to accurately determine the mechanism of the instability. As mentioned before, the relative fluctuation level is maximum at the edge of the plasma column where the density sharply decreases. The frequency of the unstable mode is close to the ion cyclotron frequency and the frequency increases when the potential of the anode of the source chamber is decreased. As explained hereupon, the flux of the ionizing electrons entering the cylindrical magnetized target plasma is larger when the anode potential inside the source chamber is decreased. The subsequent change in the plasma potential on the axis of the magnetized plasma column leads to the enhancement of the radial electric field. It is important to note that the radial electric field is inward-directed. This is of major importance in the investigation of the destabilization mechanism. During the experiments, the radial scan of a swept Langmuir probe gives the mean plasma potential profile and this in return gives an estimation of the profile of the mean radial electric field. A parabolic radial profile of the plasma potential is most often recorded near the center of the plasma column giving rise to a static inward-directed radial electric field with a linear radial profile. This leads to a rigid-body rotation of the magnetized plasma column, at least in the central part. It is important to note that this experiment is a complex multi-parameters experiment. For instance, when investigating the evolution of the unstable frequency with changing the pressure, it is of major importance to monitor the evolution of the other parameters e.g. the density gradient length and the radial electric field during the variation of the pressure. 

In the experiments reported here, two crucial parameters for the frequency selection are identified : 1- when the gas pressure is increased, the frequency of the instability always decreases; 2- when the B-field is changed, the unstable frequency varies non-monotonically.\\

\begin{figure}
	\includegraphics[scale=0.35]{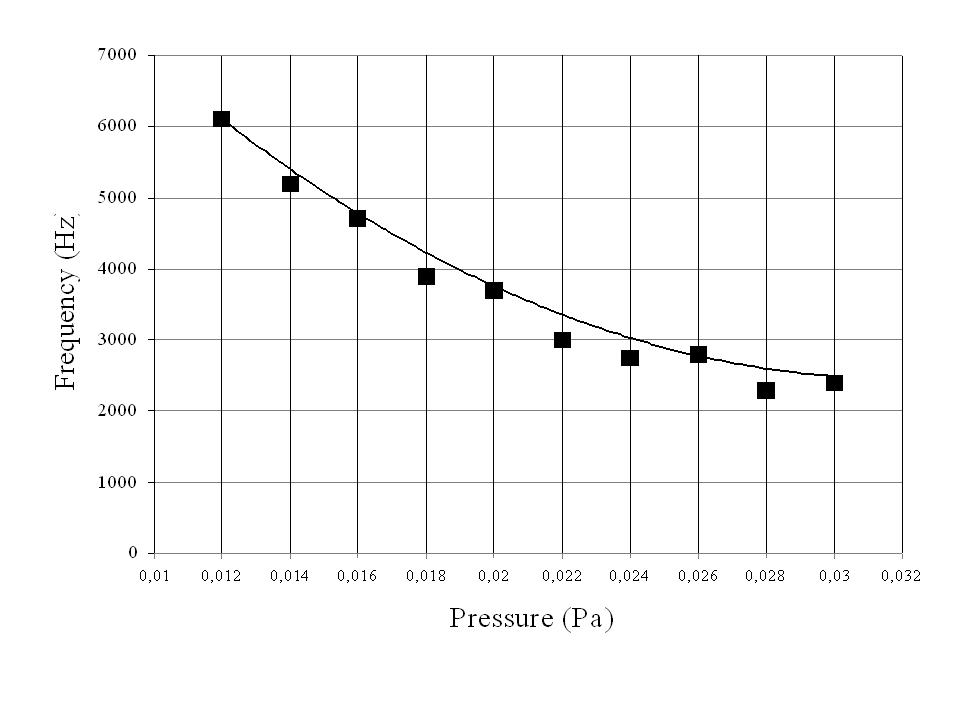}
	\caption{\label{fig:simon-fig3}  Unstable frequency in the edge of the magnetized plasma column (squares) at B = 8 mT when the pressure is increased from 0.01 Pa to 0.03 Pa. The line displays the theoretical evolution of the rotation frequency.
	}
	\end{figure}
Investigating first the effect of the pressure parameter, the unstable frequency is recorded when the gas pressure is changed over the range 0.01 Pa to 0.03 Pa at a magnetic field B = 8 mT. The recorded evolution is depicted in Fig. 3 (squares). We note that the unstable frequency is roughly inversely proportional to the pressure. This will be compared to the theory in the next section. These measurements show that the collisionality (ions-neutral collisions) is clearly a major parameter in this experiment. Note that the evolution of the electron temperature changing the pressure cannot be invoked for the decrease of the unstable frequency. In fact, in the range of pressure investigated in this work, the electron temperature decreases from 4 eV to 3 eV, inducing a relatively small change of the ion-acoustic velocity (about 15\%) that cannot explain the recorded fast decrease of the unstable frequency when the pressure is increased.\\

    In the second set of experiments, we investigate the influence of the second crucial parameter, namely the magnetic field strength. We observe that the increase of the B-field leads first to an increase of the unstable frequency when the B-field is low and then to a decrease of the frequency at higher B-field. The evolution of the frequency has been recorded at a fixed pressure 0.02 Pa in argon changing the B-field from 5 mT to 20 mT. The results are depicted in Fig. 4 that displays the unstable frequency (triangles) versus the magnetic field strength. The unstable frequency is maximum at B = 9 mT. At higher magnetic field strength, the unstable frequency progressively change for a 1/B decrease. In Fig. 4, the fluctuation level is displayed during the B-field evolution (dots). The maximum of the fluctuation level is obtained on a B-field range corresponding to the maximum values of the unstable frequency. As will be explained hereafter when detailing the theory, we argue that the maximum of the unstable frequency and its highest level are obtained when the ion mean free path is equal to the ion cyclotron Larmor radius. In other words, it seems that the parameter $\nu_{in}$/$\Omega_{ci}$, where $\nu_{in}$ is the global ion-neutral collision frequency and $\Omega_{ci}$ the ion cyclotron angular frequency, is an important parameter in the selection of the unstable frequency.\\

\begin{figure}
	\includegraphics[scale=0.35]{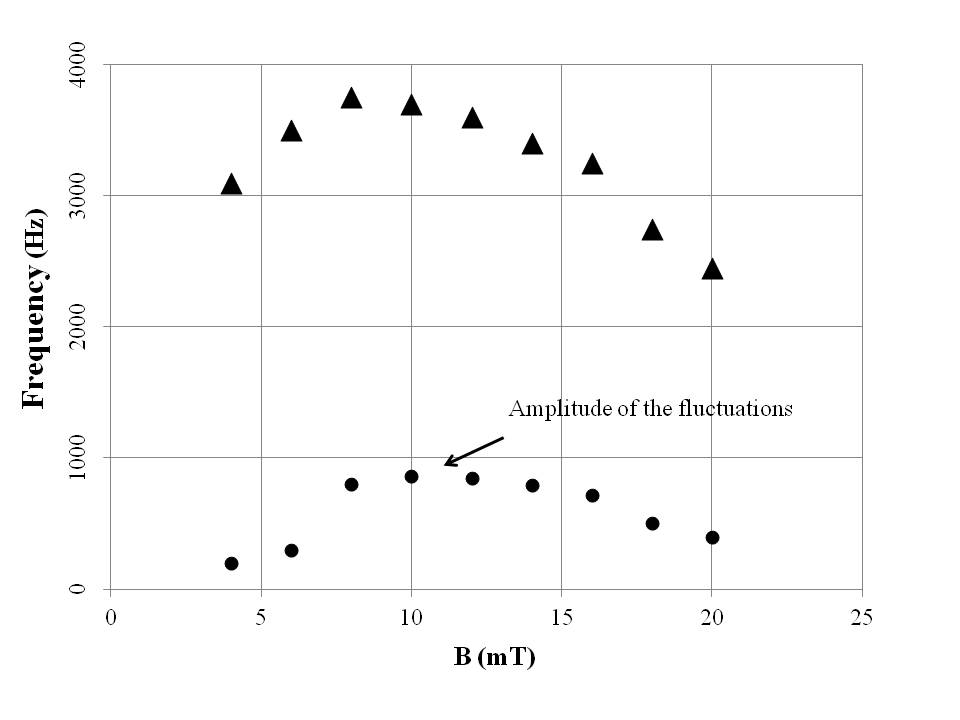}
	\caption{\label{fig:simon-fig4} Evolution of the unstable frequency (triangles) and the fluctuation level (dots) at r = 6 cm with increasing the magnetic field.
	}
	\end{figure} 
    
\section{Theoretical analysis}
    
Considering the theoretical analysis, it is well known that the transverse diffusion coefficient and the transverse mobility in a weakly-ionized magnetized plasma are dependent on the collision time between ions and neutrals. Analyzing the mobility and the diffusion of ions in a weakly ionized magnetized plasma with ion cyclotron angular frequency $\Omega_{ci}$ and ion-neutral collision time $\tau_{in}$, with the parallel mobility $\mu_0$ and the parallel diffusion coefficient $D_0$, the perpendicular mobility and perpendicular diffusion coefficient for ions are expressed as :
\begin{equation} \label{eq:perp-equation}
\mu_\perp= \mu_0 (1/ 1+ \Omega_{ci}^2. \tau_{in}^2 ) \hspace{5 mm } and \hspace{5 mm } D_\perp= D_0 (1/ 1+ \Omega_{ci}^2. \tau_{in}^2 )
\end{equation}
As a consequence, the \textbf{E}\textbf{x}\textbf{B} velocity of the ions taking into account the collisions is expressed as \cite{chen} :
\begin{equation} \label{eq:drift-equation}
V_E =   (E/B) / (1+\nu_{in}^2/\Omega_{ci}^2)
\end{equation}
In this analysis, the collision time include the global ion-neutral collision time (elastic collision time and charge exchange collision time). As a consequence, the drifting ions experience a collisional drag and the \textbf{E}\textbf{x}\textbf{B} ion drift is slow compared to the electron drift. For some plasma parameters, the reduced collisionality factor $\nu_{in}$/$\Omega_{ci}$ is dramatically changing the electric convection of the particles. Considering the instability mechanism in the case of a slab model, if a positive density disturbance is created at the edge of the plasma, a space charge separation is built inside the disturbance by the difference in the azimuthal electric drift velocity of positive and negative charges. The induced transverse electric field induced by the charge separation is perpendicular to the inward-directed electric field and it can amplify the initial perturbation because the plasma will be locally displaced at a larger radial position. The condition for amplification is that the density gradient and the electric field must be oriented in the same direction, as explained in the seminal paper by Simon (1963)\cite{simon} and in a following paper by Hoh (1963).\cite{hoh} 


The mechanism of the slow ion drift instability has been analytically studied in the case of a high temperature plasma including finite Larmor radius effects.\cite{rosen} The nonlinear theoretical analysis has been detailed later.\cite{kim} This instability has also been identified as an important mechanism in a toroidal laboratory plasma.\cite{kaur} The difference in velocity drift of positive and negative species has been shown responsible for instabilities in non-collisional plasmas exhibiting large ion radius effects inside a cylindrical magnetized plasma\cite{sakawa}, that is the so-called Modified Simon-Hoh high frequency instability.
 It is important to have in mind that the ion diamagnetic drift that experience also a collisional drag is negligible in our experiment. The low ion temperature (about 0.1 eV) and the gradient length (2 to 4 cm) leads to an ion diamagnetic drift (opposite to the \textbf{E}\textbf{x}\textbf{B} drift) that is least ten times lower that the electric drift.

\section{Comparison with experiment}
  
  We now compare the measurements to the theoretical analysis of the slow ion drift instability. We investigate first the theoretical variation of the unstable frequency when the gas pressure is changed. The B-field value in the calculation is 8 mT. The main parameter in the physical mechanism described here is the global ion-neutral collision frequency. In fact, the choice of the numerical value for the collision frequency is rather complex. We refer to the literature detailing the collisions between ions and neutral, including elastic collisions\cite{phelps} and charge exchange collision\cite{maiorov}, giving the cross-section for the interaction in the range 0.8 $10^{-17} m^{2}$ and 1.2 $10^{-17} m^{2}$. At the pressure P = 0.02 Pa in argon with ion temperature $T_i$= 0.1 eV, we have chosen $\nu_{in}$ = $10^{4}$Hz for the calculation. This corresponds to a mean free path of 5 cm at the ion thermal velocity 490 m/s. This value for the collision frequency is similar to the value used in previous numerical simulation of the unstable magnetized plasma column \cite{naulin} where the normalized collision  frequency was $\nu_{in}$/$\Omega_{ci}$ = 0.06 at B = 60 mT though the charge exchange collisions were not taken into account. In the theoretical evaluation of the E\textbf{x}B ion drift velocity when the pressure is varied, we use that reference value of the global ion-neutral collision time $\tau_{in}$ = $10^{-4}$s at P = 0.02 Pa with $\Omega_{ci}$ = 1.9 $10^{4}$ rad/s at B = 8 mT. \par

 In order to compare the experimental results to the theory, we investigate first the variation of the unstable frequency when the gas pressure is changed. The B-field value in the calculation is 8 mT. Assuming that the angular frequency of the drifting ions determines the detected unstable frequency at the radial position r= 6.5 cm (assuming m=1 mode), the theoretical results are superimposed in Fig. 3. The solid line gives the frequency obtained with an azimuthal velocity equal to the slow ion drift velocity with a collision time changing from 0.2 $10^{-4}$s to 1.8 $10^{-4}$s when the pressure is changed from 0.012 Pa to 0.03 Pa. A good correlation with the experimental points is obtained. We conclude that the measured unstable frequency and its evolution when the pressure is increased are in accordance with the theory of the slow ion drift instability when the most unstable frequency is selected by the m=1 mode and by the slow collisional E\textbf{x}B ion drift velocity.

The theoretical investigation of this physical situation was first detailed in the paper by Simon\cite{simon} describing a flute-type Rayleigh-Taylor instability predicted in the case of a weakly ionized magnetized plasma column with an inward radial electric field and a Gaussian profile of the density. In this physical situation, the calculation indicates that the most unstable frequency is maximum when the parameter $\nu_{in}$/$\Omega_{ci}$ is close to unity. The details of the calculation and theoretical arguments are been revisited recently.\cite{steve} 

  The second important parameter determining the value of the unstable frequency is the magnetic field strength. In order to compare  the theoretical evolution of the unstable frequency to the measurements when the B-field is increased, the evolution of the slow ion drift velocity given by Eq. \eqref{eq:drift-equation} is computed between B = 5 mT and B = 20 mT at a pressure 0.02 Pa.

\begin{figure}
	\includegraphics[scale=0.3]{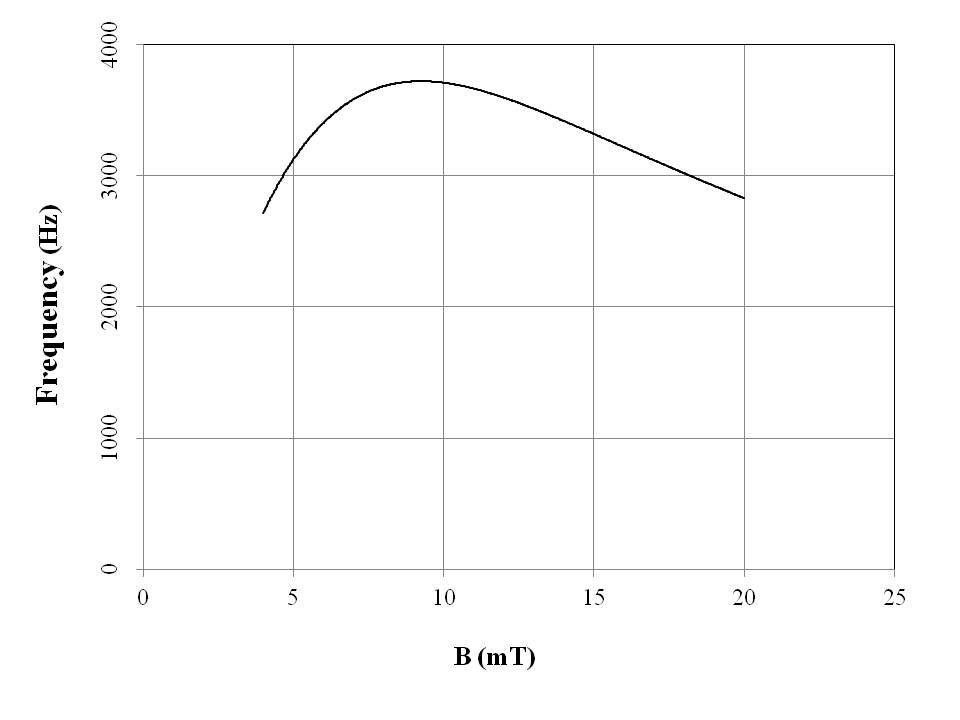}
	\caption{\label{fig:simon-fig5} Theoretical evolution of the angular rotation frequency of the drifting ions with parameters: radial position r = 6.5 cm, radial electric field 20 V/m, ion-neutral collision frequency 18 kHz. A lower collision frequency, for instance 11 kHz, would give B = 6 mT for the value of the B-field corresponding to the most unstable frequency.}
	\end{figure}
    
The evolution of the angular frequency of the magnetized ions drifting at the slow collisional E\textbf{x}B drift velocity is displayed in Fig. 5. The  global collision frequency has been adjusted in order to get a maximum of the unstable frequency at 9 mT,
that leads to $\nu_{in}$ = 1.8 $10^{4}$ Hz. This value leads to a mean free path of 2.7 cm (ion temperature $T_i$= 0.1 eV). A rigid-body rotation induced by a radial electric field of 20 V/m at r = 6.5 cm is assumed in the calculation. The non-monotonous variation of the angular frequency displayed in Fig. 5 is similar to the experimental results depicted in Fig. 4. At the frequency maximum (B = 9 mT), the ion Larmor radius is 2.27 cm i.e. a value very close to the mean free path.
 It is important to note that this evolution of the unstable frequency when the B-field is increased is very sensitive to the ion-neutral collision time. For instance, choosing  $\nu_{in}$ = 1.1 $10^{4}$ Hz would give a maximum of the unstable frequency at B = 6 mT.
 
 \section{Conclusion}
  The low-frequency flute-type instability observed in the magnetized plasma column produced in our laboratory device has been investigated taking into account the low magnetization of the ions leading to a relatively large ion Larmor radius of argon ions. The cyclotronic movement of the ions moving at thermal velocity leads to a radial excursion comparable to the radius of the plasma column. Moreover the gas pressure used in the experiment leads to a high probability for created ions to be deflected or neutralized after a few centimeters. The existence of a radial electric field directed toward the axis of the plasma column induces a global rotation of the plasma ions. The rotation is slowed down by the collisions with neutrals and in return induces the low-frequency instability of the magnetized plasma column. This experimental situation has been compared successfully to the description of the slow electric drift instability described first by A. Simon\cite{simon} in the early 60's.
The measurements indicate a good agreement with the theory, especially about the selection of the most unstable frequency when the mean free-path of the ions is close to ion Larmor radius. In this situation, an enhancement of the transport across the magnetic field is predicted.

 In conclusion, the parameter $\nu_{in}$/$\Omega_{ci}$, where $\nu_{in}$ is the global ion-neutral collision frequency and $\Omega_{ci}$ the ion cyclotron angular frequency, is an important parameter for the stability of a poorly magnetized collisional plasma.

\begin{acknowledgments}
We are grateful to Prof. F. F. Chen for suggestions about the Simon-Hoh instability and to Prof. G. Tynan for proposing this mechanism when visiting our experimental set-up.\\
This paper is dedicated to the memory of Dr. Steve Jaeger (1976-2014), a dear friend and an inspiring collaborator.
\end{acknowledgments}

\end{document}